\pdfoutput=1 

\documentclass[3p,times]{elsarticle}

\usepackage{ecrc}

\usepackage{epstopdf}

\volume{00}

\firstpage{1}

\journalname{Nuclear Physics A}

\runauth{M. Floris}

\jid{nupha}

\jnltitlelogo{Nuclear Physics A}

\usepackage{graphicx}
\usepackage{amsmath,amssymb}
\usepackage{mathrsfs}

\usepackage{lineno}

\usepackage{color}

\newcommand{\kzero}        {\ensuremath{{\rm K}^{0}_{S}}}
\newcommand{\kstar}        {\ensuremath{{\rm K}^{*}}}
\newcommand{\He}           {\ensuremath{^{3}{\rm He}}}
\newcommand{\LH}           {\ensuremath{^{3}_{\Lambda}{\rm H}}}

\newcommand{\PbPb}         {\mbox{Pb--Pb}}
\newcommand{\pPb}          {\mbox{p--Pb}}
\newcommand{\AuAu}         {\mbox{Au--Au}}

\newcommand{\dNdy}         {\ensuremath{\mathrm{d}N/\mathrm{d}y}}

\newcommand{\s}            {\ensuremath{\sqrt{s}}}

\newcommand{\ppi}          {\ensuremath{{\rm p}/\pi}}
\newcommand{\kpi}          {\ensuremath{{\rm K}/\pi}}

\newcommand{\snn}          {\ensuremath{\sqrt{s_{\rm NN}}}}

\newcommand{\Tch}          {\ensuremath{{T}_{\rm ch}}}

\newcommand{\muB}          {\ensuremath{\mu_{B}}}

\newcommand{\gs}           {\ensuremath{\gamma_{s}}}
\newcommand{\gq}           {\ensuremath{\gamma_{q}}}
\newcommand{\gc}           {\ensuremath{\gamma_{c}}}
\newcommand{\chindf}       {\ensuremath{\chi^{2}/{\rm NDF}}}

\newcommand{\etalab}       {\ensuremath{\eta_{{\rm lab}}}}

\begin{document}

\begin{frontmatter}



\title{Hadron yields and the phase diagram of strongly interacting matter}

\author{M Floris}
\address{CERN, CH-1211 Geneva 23, Switzerland}




\begin{abstract}
  This paper presents a brief review of the interpretation of
  measurements of hadron yields in hadronic interactions within the
  framework of thermal models, over a broad energy range (from SIS to
  LHC energies, $\snn\simeq 2.5$~GeV -- 5 TeV). Recent experimental
  results and theoretical developments are reported, with an emphasis
  on topics discussed during the Quark Matter 2014 conference.

\end{abstract}


\end{frontmatter}


\section{Introduction}
\label{intro}

It has been known since more than two decades that hadrons in high
energy interactions are produced in approximate thermal and chemical
equilibrium~\cite{Becattini:2009fv,Cleymans:1998fq,BraunMunzinger:2003zd,Petran:2013lja}.
The relative abundances of light flavor hadrons are determined by a
few thermal parameters and, in the simplest case, approximately
proportional to the Boltzmann factors.  This is true with the only
exception of strange particles, which deviate from the expected
equilibrium abundance by a factor which depends on the strangeness
content of the particle~\cite{Rafelski:1982pu,Tounsi:2001ck}.  In
central heavy ion collisions~\cite{BraunMunzinger:2003zd}, strange
particles were however found to follow the expected equilibrium
distribution\footnote{If $4\pi$ yields are considered instead of
  mid-rapidity \dNdy, a strangeness under-saturation factor of about
  0.75 is still required at the SPS. Also note that the usage of
  midrapidity \dNdy\ at low energy is questionable, see for instance
  the discussion in ~\cite{Becattini:2009fv,BraunMunzinger:2003zd}.}.

Measurement at different \s\ revealed that the thermal fit parameters follow a smooth
curve in the plane of temperature $\Tch$ and baryo-chemical potential
$\muB$, the so-called ``hadron freeze-out
curve''~\cite{Cleymans:1998fq,Cleymans:2005xv}. For very high energy
collisions, $\muB\to0$ and the temperature extracted from
a thermal analysis of the data ($\Tch\sim 150-170 ~\mathrm{MeV}$) is
found to be very close to the critical (crossover) temperature
estimated in lattice QCD, $T_\mathrm{c} \in (143,171)~\mathrm{MeV}$~\cite{Borsanyi:2010bp}.  Despite these
observations, the profound meaning of the freeze-out
curve~\cite{Cleymans:2005xv,Stock:1999hm,BraunMunzinger:2001mh,BraunMunzinger:2003zz,Andronic:2009gj} remains unclear and several key questions remain unanswered:
\begin{itemize}
\item What is the relation of the chemical freeze-out temperature to the QCD critical temperature?
\item How is the equilibrium reached?
\item What physics mechanisms drive the hadron freeze-out curve?
\end{itemize}

Recently, higher precision measurements revealed unexpected deviations from
the thermal model expectations.  The statistical model is an effective
model (see e.g.\ the discussion in~\cite{Becattini:2010sk}) and the
small deviations from the equilibrium picture may simply indicate that
the precision of the data has become sufficient to reveal its
limitations. Their study represents an opportunity for a
better understanding of the underlying physics processes.

A qualitative summary of the parameters which control a thermal fit is
given below for the non-expert reader. The interested reader should
refer e.g.\
to~\cite{Becattini:2009fv,BraunMunzinger:2003zd,Petran:2013lja} and
references therein for a rigorous discussion.

The main parameters, describing particle production in equilibrium are:
\begin{itemize}
\item The temperature \Tch\ (constrained by ratios of particles with a
  large mass difference);
\item The baryochemical potential \muB\ (constrained by
  anti-baryon/baryon ratios, at LHC it is found
  $\muB\sim0$);
\item The volume V, which acts as a normalization parameter
  (constrained by the most precisely measured species, typically pions).
\end{itemize}

\begin{figure}[tp]
  \centering
  \includegraphics*[width=0.58\textwidth]{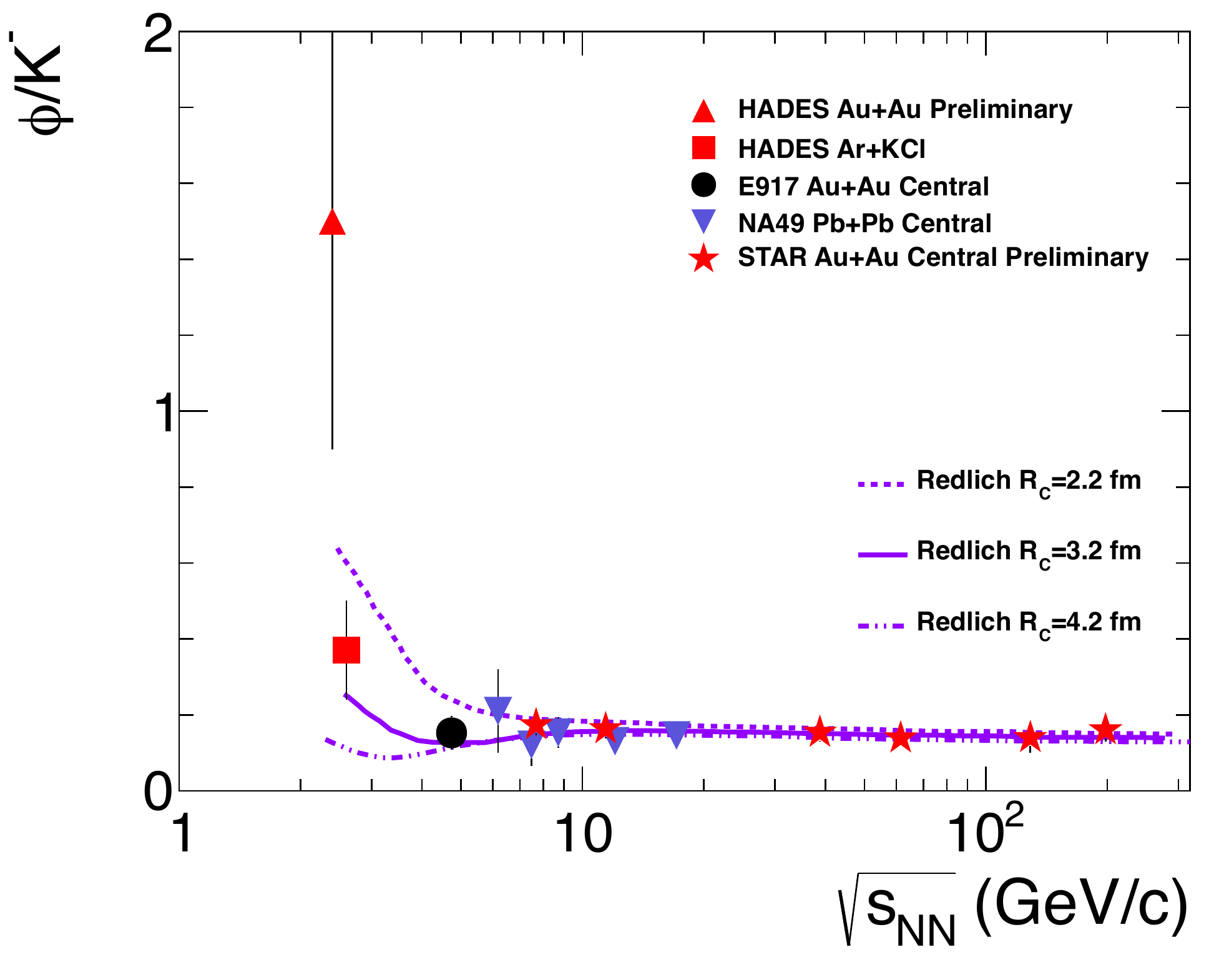}
  \caption{$\phi/\mathrm{K}^{-}$ ratio as a function of \s\ in AA collisions~\cite{Lorenz:QM14}.}
  \label{fig:phi-vs-roots}
\end{figure}

Deviations from (grand canonical) equilibrium can be incorporated
through empirical under(over)-saturation parameters for strange, charm
or light quarks (\gs, \gc\ and \gq).  The need for \gs\ has already
been discussed above. Another approach consist in the
implementation of the ``canonical suppression'' mechanism (i.e.\
strangeness has to be conserved exactly and not on
average)~\cite{Tounsi:2001ck} on a smaller volume than the overall
size of the system, determined by a ``canonical radius'' parameter,
$R_C$.  The parameter \gc\ is introduced because charm can only be
created in the initial phases of the collisions (it is too heavy to be
created thermally)~\cite{BraunMunzinger:2000px} and it is thus expected
to be significantly out of equilibrium.  While the usage of \gs\ and
\gc\ is common to most implementations of the statistical
model~\cite{BraunMunzinger:2003zd,Wheaton:2004qb,Becattini:2009sc},
\gq\ is only found in the non-equilibrium model
SHARE~\cite{Petran:2013lja}. The physical picture in this model is
that of an expanding, super-cooled quark-gluon plasma which undergoes
a sudden hadronization without further re-interactions.  The thermal
parameters of the quark-gluon plasma are hence frozen, leading to
out-of-equilibrium hadron abundances. From the point of view of the
fit, \gq\ allows the relative abundance of mesons and baryons to vary (as it
is determined by the number of valence light quarks).

The selection of results presented in this paper reflects the author's
bias, especially towards results presented during the Quark Matter
2014 conference. Due to space limitations, the list of references is
also incomplete and subject to similar biases: preference is given
to reviews and recent papers over the original literature.

\section{Results at low energy}
\label{sec:low-energy}

The HADES collaborations published results on the thermal description
of eight particle species ($\pi^{-}$, $\Lambda$, $\mathrm{K}^{+}$,
\kzero, $\phi$, $\Xi^{-}$, $\Sigma^{+-}$) in Ar-KCl collisions
(approximately 40 nucleons impinging on 40 nucleons) at
\snn~=~2.61~GeV~\cite{Agakishiev:2010rs}. These results have been
recently complemented by a new fit of four particle ratios
($\pi^{-}/p$, $\kzero/\Lambda$, $\mathrm{K}^-/\mathrm{K}^{+}$,
$\phi/\mathrm{K^{-}}$) measured in Au-Au collisions at \snn~=~2.4~GeV,
presented for the first time at this conference~\cite{Lorenz:QM14}. In
general, the agreement of these very low energy results with the
thermal model calculation is surprisingly good, except for the
$\Xi^{-}$ baryon which is underestimated by about an order of
magnitude in the model~\cite{Agakishiev:2010rs}.
The $\phi/\mathrm{K}^{-}$ ratio shows a hint of an increase at low
\snn\ (Fig.~\ref{fig:phi-vs-roots}). This is consistent with the
expectation from the canonical suppression mechanism: the $\phi$ meson
(being a hidden strangeness particle) is not suppressed, while the K
is, leading to an increase in the $\phi/\mathrm{K}^{-}$ ratio as the
energy decreases.  This observation is in contrast with results at SPS
energies in pp and peripheral \PbPb\ collisions, where the canonical
suppression mechanism did not describe the $\phi$ data (while the
small colliding system size would suggest that it plays an even more
important role)~\cite{Kraus:2008fh,Becattini:2008yn}.  It is also
interesting to notice that the $\phi/\mathrm{K}^{-}$ ratio is $\sim 1$
at $\snn\sim2.5$~GeV, suggesting that about half of the
$\mathrm{K}^{-}$ originate from $\phi$ decays.

\section{pp Collisions at the LHC}
\label{sec:pp-collisions-lhc}

\begin{figure}[tp]
  \centering
  \includegraphics*[width=0.7\textwidth]{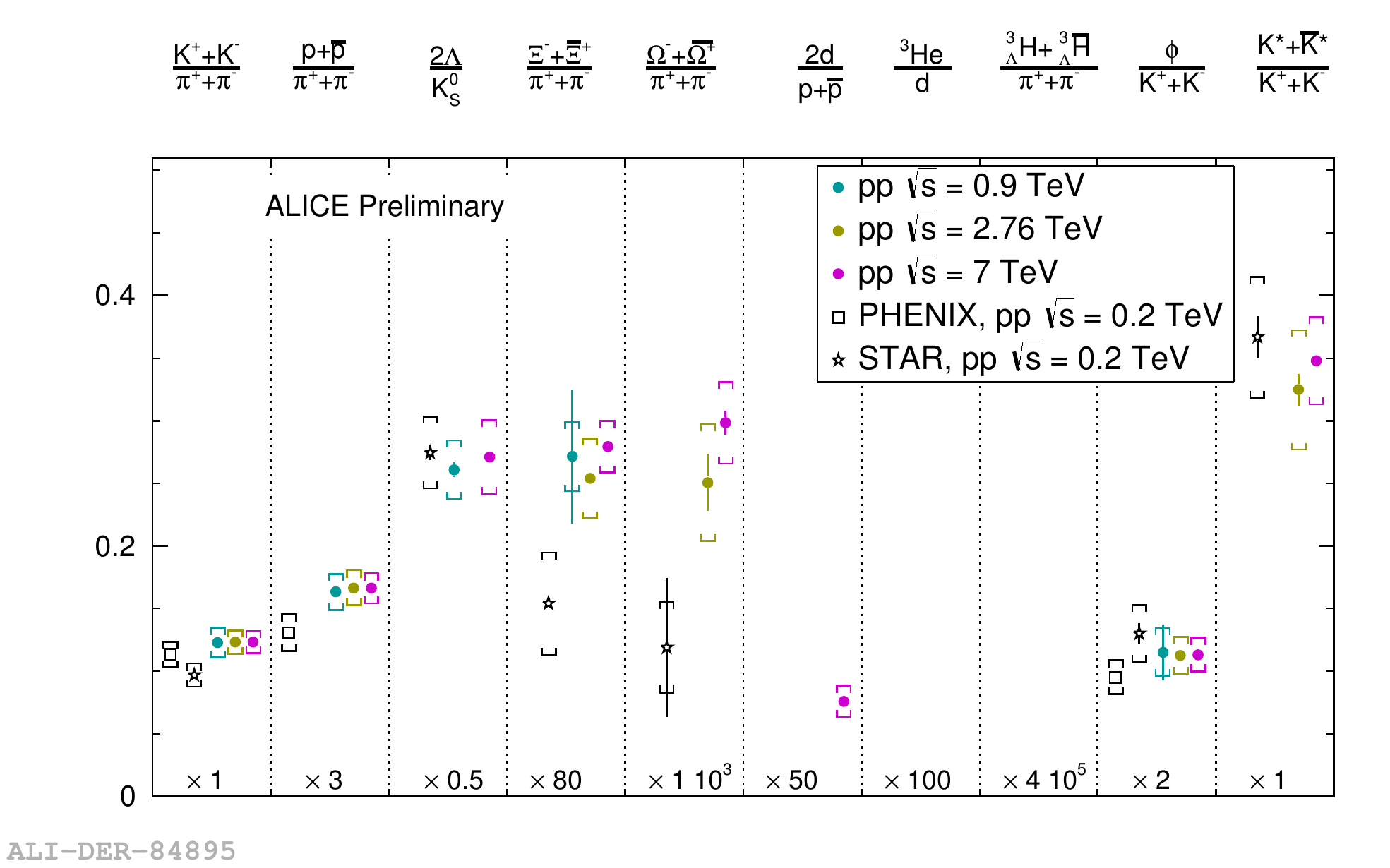}
  \caption{Particle ratios in pp collisions for different \s~(the \He\
    and \LH\ are not yet measured in pp collisions at the LHC. The
    plot shows two empty columns to allow for an easier visual
    comparison with the \PbPb\ ratios, shown in
    Fig.~\ref{fig:ratios-vs-s-AA}).}
  \label{fig:pp-ratios-vs-roots}
\end{figure}

The ALICE collaboration measured the production yield of several
particle species in pp collisions at \s~=~0.9, 2.76, and 7~TeV. No
changes in particle ratios are observed for $\s\gtrsim1~\mathrm{TeV}$,
as shown in Fig.~\ref{fig:pp-ratios-vs-roots}. On the other hand, if
the measurements are compared to results at top RHIC energy
(\s~=~200~GeV), significant changes are observed. Most notably, the
production of (multi)strange particles relative to pions is seen to
increase and then saturate at LHC energies. It was suggested that the
grand-canonical limit could be reached in pp collisions at the
LHC~\cite{Becattini:2009ee}, and this observation would seem to
support this idea. However, an actual grand-canonical fit of pp data
at the LHC yields a poor quality, with a $\chindf > 7$ and an increase
with system size of the ratio of multistrange baryons to pions is
still seen at LHC energies (Fig.~\ref{fig:ratios-vs-systemsize}).  The
absolute deviations data/model are smaller than 20-30\%.  As a final
remark, we notice a hint for an increase of the \ppi\ ratio with
energy (while an opposite trend is suggested by AA data,
sec.~\ref{sec:heavy-ion-collisions}).

\section{System size dependence at the LHC}
\label{sec:system-size-lhc}

Many new results on identified particle production as a function of
charged multiplicity in \pPb\ collisions at \snn~=~5.02~TeV were
reported by the ALICE collaboration. These studies play an important
role in the understanding of hadron production, as they bridge pp and
\PbPb\ collisions in terms of multiplicity.  In general, the selection
of events based on charged particle multiplicity in small systems can
introduce non-trivial dynamical biases~\cite{Toia:QM14,Adare:2013nff}, and results
can depend on the choice of the multiplicity estimator.  All the results
presented in this section were binned using quantiles in the amplitude
of the ALICE V0A detector, a forward hodoscope covering $2.8 <
\etalab\ < 5.1$ in the lead-going direction~\cite{Abelev:2014ffa}.
Most of the ratios which change between pp and \PbPb\ collisions
attain an intermediate value in \pPb\ collisions.

\begin{figure}[tp]
  \centering
  \includegraphics*[width=0.7\textwidth]{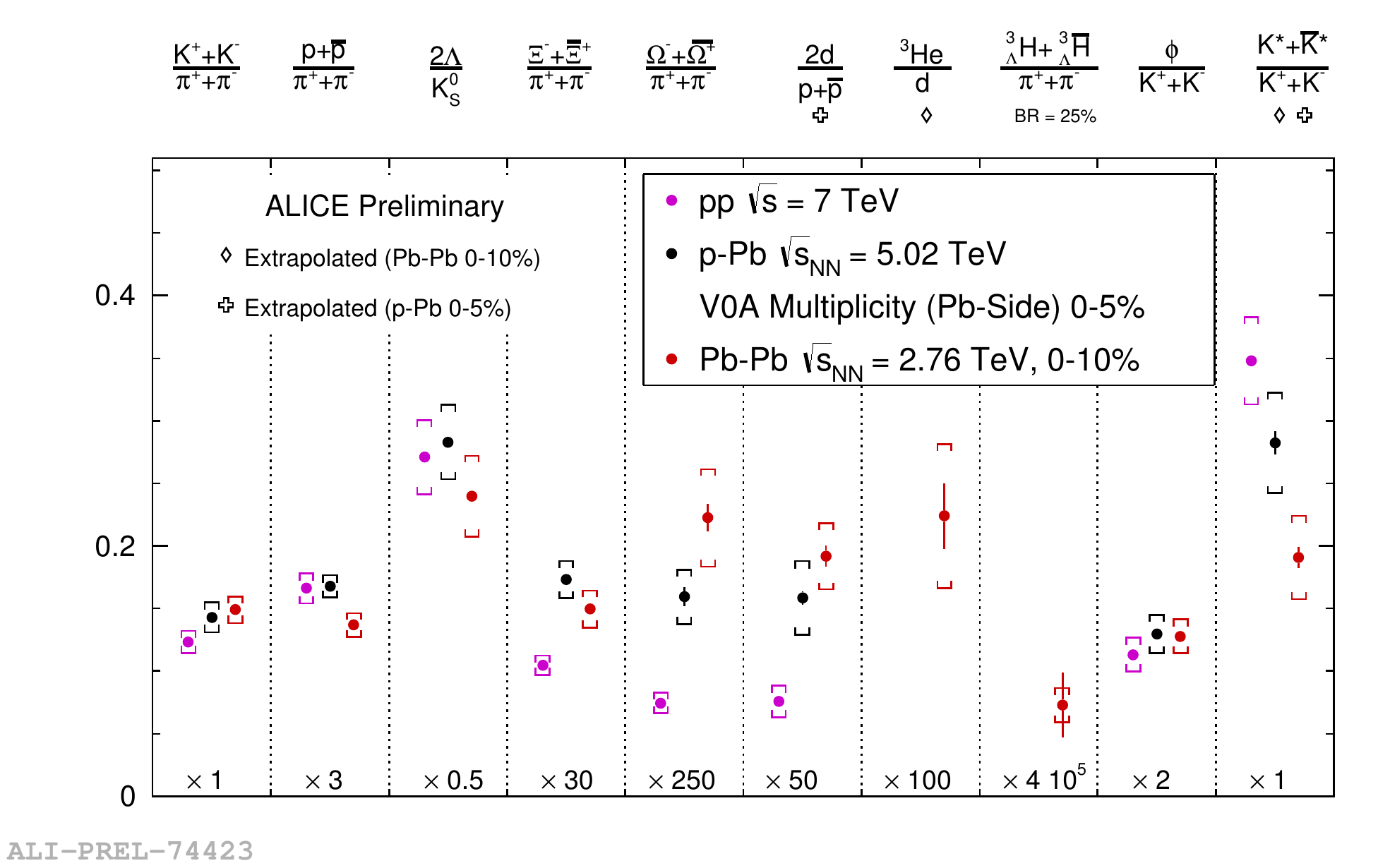}
  \caption{Particle ratios in pp, \pPb, and \PbPb\ collisions at LHC energies.}
  \label{fig:ratios-vs-systemsize}
\end{figure}

(Multi)strange hadrons are seen to increase as a function of
multiplicity in \pPb\ collisions, reaching almost grand-canonical
values (with the same temperature $\Tch\simeq156~\mathrm{MeV}$
observed in \PbPb\ collisions), see
Fig.~\ref{fig:ratios-vs-systemsize} and \cite{Alexandre:QM14}. There
is, however, a small tension between the yields of the $\Xi$ and
$\Omega$ baryons normalized to pions: the former is higher than the
corresponding \PbPb\ ratio for the highest multiplicity \pPb\
collisions, but the latter is below the \PbPb\ value. While none of
the two is individually significant within uncertainties, their
combined effect leads to a large \chindf\ ($\sim 5$) when a
grand-canonical fit of high multiplicity \pPb\ results is
performed. The origin of this tension still needs to be
clarified.

The \kstar/K ratio is seen to decrease with increasing multiplicity,
as expected in case the \kstar\ decay products undergo elastic
scattering in the hadronic medium~\cite{Bellini:QM14}.  The d/p ratio
is observed to increase with increasing multiplicity in \pPb\
collisions, while it is flat as a function of centrality in \PbPb\
collisions~\cite{Martin:QM14}. This observation constrains the
production mechanism of deuterons (thermal production or coalescence
models). At this conference, the STAR collaboration reported results
on $v_2$ of nuclei at RHIC energies, which would support coalescence
as the dominant production mechanism~\cite{Haque:QM14}. Detailed
theoretical modeling as a function of \snn\ would be needed to draw
any firm conclusion.
 
Finally, we notice that there is a hint of a decrease of the
baryon/pion ratios ($p/\pi$ and $\Lambda/\pi$), which would be
consistent with the baryon annihilation mechanism proposed to explain
particle ratios in central \PbPb\ collisions discussed in the next
section~\cite{Becattini:2014hla}.

\section{Heavy Ion Collisions at High Energy}
\label{sec:heavy-ion-collisions}

\begin{figure}[tp]
  \centering
  \includegraphics*[width=0.7\textwidth]{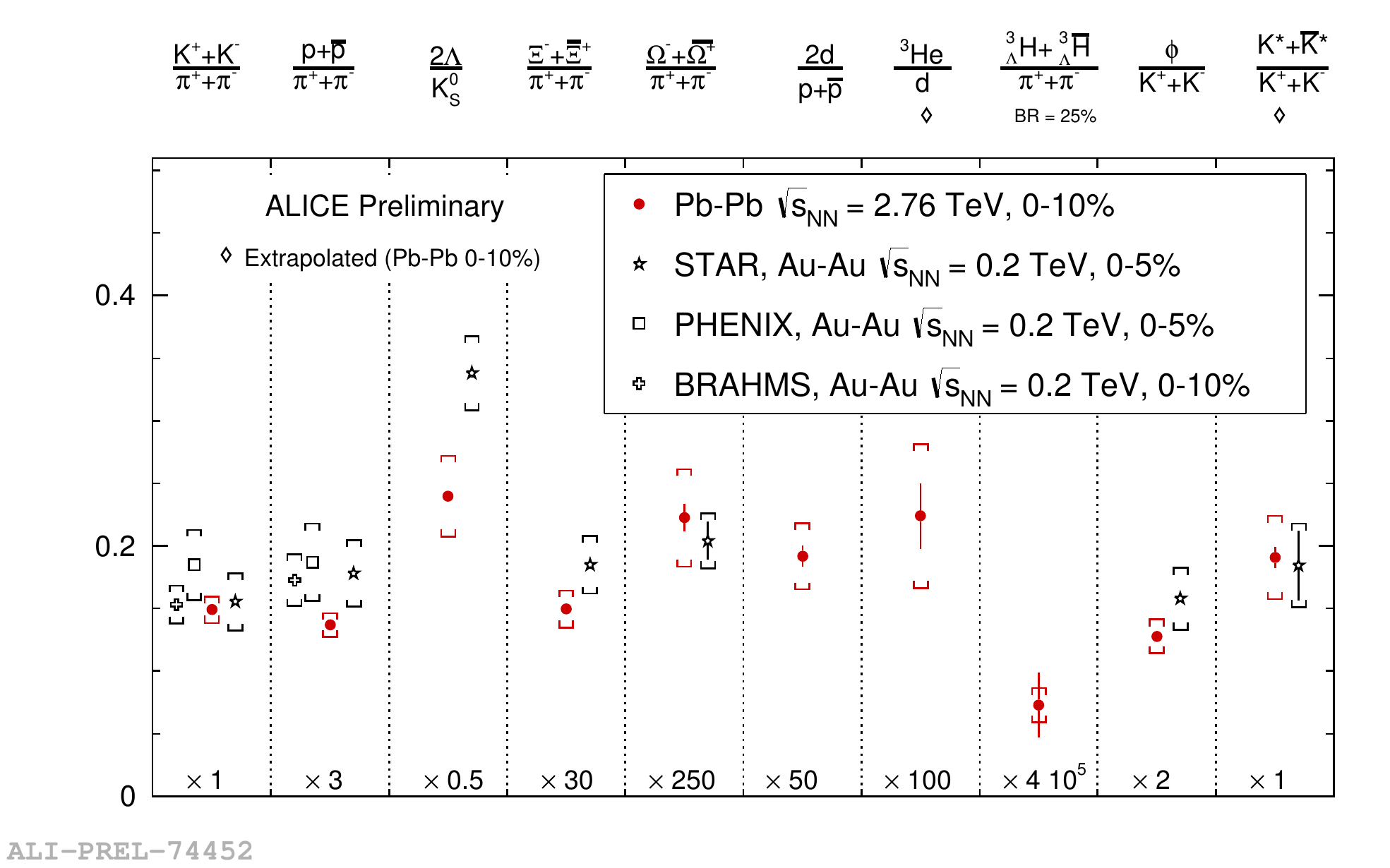}
  \caption{Particle ratios at top RHIC and at LHC energy in AA
    collisions. In this figure, STAR protons have been corrected for
    feed-down based on the thermal model
    estimate.}  \label{fig:ratios-vs-s-AA}
\end{figure}

\begin{figure}[tp]
  \centering
  \includegraphics*[width=0.7\textwidth]{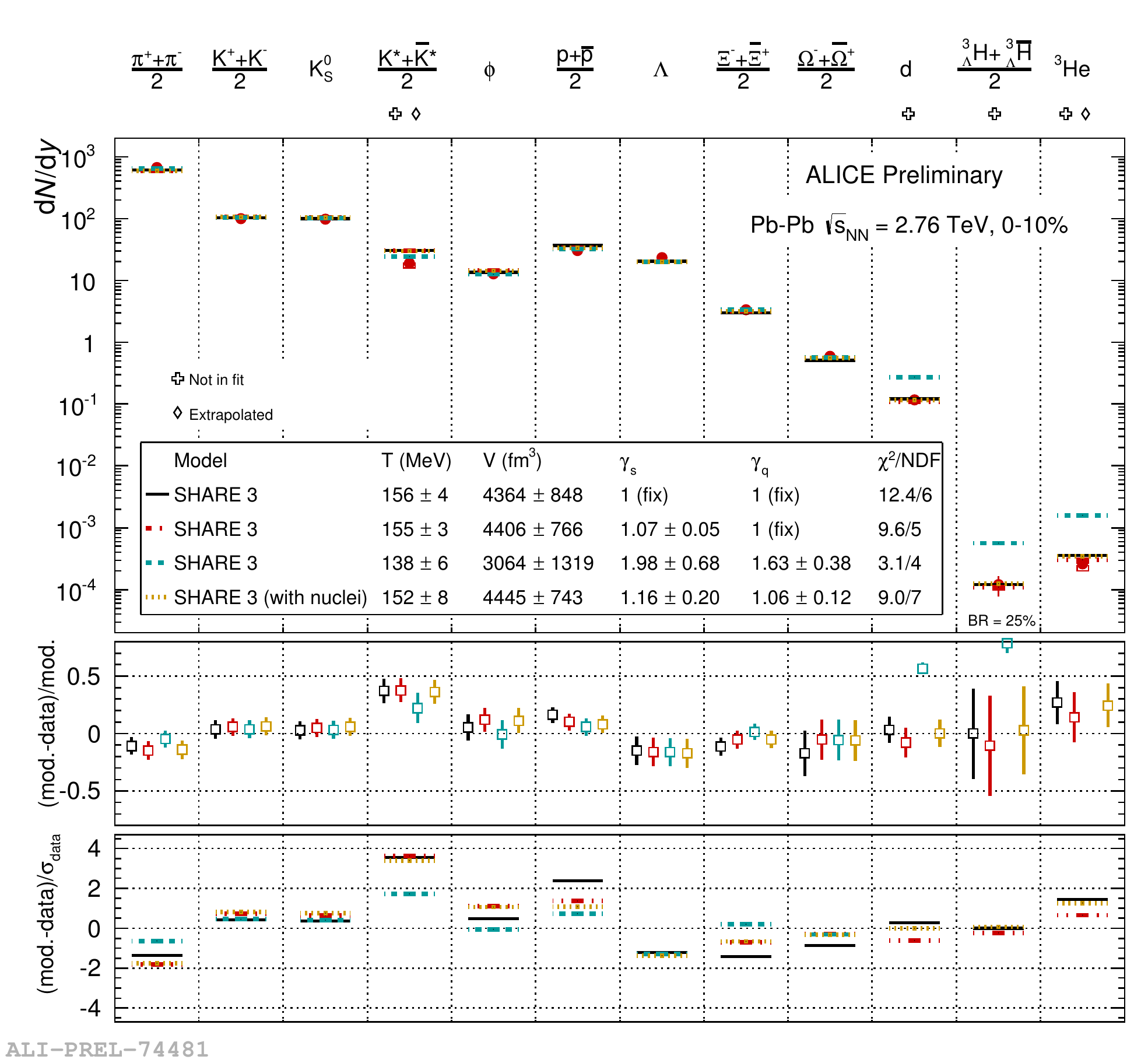}
  \caption{Thermal fit of particle yields in 0-10\% central \PbPb\ with the SHARE model for various scenarios.}
  \label{fig:share-different-fits}
\end{figure}

Central heavy ion collisions are commonly regarded as the ideal system
for the applicability of the thermal model.  Equilibrium thermal
models were found to give an excellent description of particle yields
measured in \AuAu\ collisions at top RHIC energy (\snn~=~200~GeV,
$\chindf\simeq 1$ when analyzing data from the same experiment, $\chindf
\sim 2$ when fitting all RHIC results
simultaneously~\cite{Andronic:2008gu}).  The analysis of data as a
function of \snn, moreover, allowed to extrapolate the parameters to
the limiting case of very high energy, which is essentially reached at
the LHC~\cite{Andronic:2008gu,Cleymans:2006xj}. Therefore, the
observation of an anomalously low $\ppi = 0.046\pm 0.003$ ratio in
central collisions at the LHC (a factor $\sim 1.5$ lower then
expectations based on
\Tch~=~164~MeV)~\cite{Abelev:2012wca,Abelev:2013vea} came as a
surprise and triggered considerable theoretical discussions. A low
value of the \ppi\ ratio is naturally predicted by non-equilibrium
models, as a consequence of the lower temperature ($\Tch \sim
140$~MeV) needed to describe the data.  The predictions of the \ppi\
ratio with the preferred set of parameters in~\cite{Rafelski:2010cw}
were indeed in agreement with the data. However, with the same
parameters, the \kpi\ ratio was over-estimated by a similarly large
factor.  The ``anomaly'' with respect to the equilibrium model
expectations is now confirmed by fits which include a complete set of
particle species (Fig.~\ref{fig:share-different-fits}). It is
furthermore observed that the consistency with equilibrium
expectations (and extrapolation of RHIC results) is essentially
restored if protons and the \kstar\ resonance are excluded from the
fit~\cite{Andronic:2012dm}.

Given the tension between results at \snn~=~2.76~TeV and the
expectations based on lower energy results, it is natural to ask if
there is any change in particle ratios between RHIC and LHC energies,
irrespective of any thermal model interpretation. The results are
discussed below for 0--10\% central collisions at the
LHC~\cite{Martin:QM14,Abelev:2013vea,ABELEV:2013zaa,Abelev:2013xaa,Abelev:2014uua}
and for 0--5\% and 0--10\% results at
RHIC~\cite{Abelev:2008ab,Arsene:2005mr,Adler:2003cb,Adams:2006ke,Aggarwal:2010mt,Abelev:2008aa,Agakishiev:2011ar}.
If positive and negative charge states are averaged, the effect of the
small charge asymmetry present in mid-rapidity yields at RHIC becomes
negligible and particle ratios are expected to be the same at RHIC and
LHC\@. No strong evidence for changes is observed, but the \ppi\ and
$\Lambda/\pi$ ratios are found to be lower at the LHC by a factor
$\sim 2.3\sigma$.  More recent results from \AuAu\ collisions at
\snn~=~62.4~GeV and \snn~=~130~GeV at RHIC show a similar tension
between the measured proton yield and equilibrium thermal model
calculations~\cite{Andronic:2012dm}.

A fit of this extended set of particle species
(Fig.~\ref{fig:share-different-fits}) with an equilibrium model yields
a \chindf\ of about 2, which is better than any other colliding system
at the LHC, but slightly worse than expected from the fit quality at
RHIC\@. The largest contribution to the \chindf\ comes from the low
yield of protons relative to pions. This conclusion persists for the
three different thermal model implementations which were used by the
ALICE
collaboration~\cite{Petran:2013lja,Wheaton:2004qb,Andronic:2008gu},
indicating that the residual differences in those models (minor
difference in the hadron list, treatment of charm and of the hadron
excluded volume) have a second order effect. The temperature obtained,
$\Tch = 156 \pm 2~\mathrm{MeV}$, is lower than at RHIC.  At this
conference, the STAR collaboration reported on high-statistics
$\Omega$ measurements in central \AuAu\ and pp
collisions~\cite{Zhu:QM14}. The measured yield in central \AuAu\
collisions is consistent with the previous result, but lower.  This
lower yield would be consistent with the chemical freeze-out
temperature extracted at the LHC\@. It will be interesting to study
how this affects the thermal fits at RHIC\@.

A separate discussion should be made for (hyper)nuclei.  Having a
binding energy much smaller than the estimated temperature at chemical
freeze-out, it may be expected that their main production mechanism is
not thermal, but they are rather formed by coalescence
(sec.~\ref{sec:system-size-lhc}). Their inclusion in thermal fits can
be questionable. However, it has been argued that a thermal approach
is still applicable because hadronic interactions do not change the entropy
per baryon and the statistical hadronization and coalescence
approaches yield compatible
results~\cite{Siemens:1979dz,Stachel:2013zma}. 
The ALICE collaboration released results on the production of
deuterons, \He\ and \LH, found to be in agreement with the
equilibrium fits.

Different explanations have been proposed in the literature to explain the particle yields measured at the LHC and 
the ``anomaly'' in the \ppi\ ratio, as summarized below.

\paragraph{Incomplete Hadron List} One of the basic ingredients in the
thermal models is the list of hadrons and high mass resonances which
feed-down to the (stable) measured species. It is known that this list
is incomplete, and it was argued (see e.g.~\cite{Stachel:2013zma})
that this could explain the tension, as decays of high mass resonances
would affect pions more than protons.  Quantitative calculations were
made
in~\cite{NoronhaHostler:2007jf,Noronha-Hostler:2014usa,Noronha-Hostler:QM14},
where it was shown that reasonable assumptions on high mass resonances
based on the Hagedorn spectrum could explain the low \ppi. However,
these additional states could potentially spoil the agreement with
other particle ratios (most notably, multistrange baryons, which could
however be included as discussed~\cite{NoronhaHostler:2009cf}) and
some of the underlying assumptions of the model are not constrained by
first principles.

\paragraph{Non-Equilibrium Thermal Model}
In the framework of the non-equilibrium thermal model, as implemented
in the SHARE code~\cite{Petran:2013lja}, it is possible to find a set
of parameters which describes all hadrons except nuclei, with a very
good \chindf~\cite{Petran:2013lja,Rafelski:2014fqa}. An interesting
feature of this description is that the physical parameters of the
fireball at freeze-out (pressure, energy density and entropy density)
are rather constant as a function of energy and
centrality~\cite{Rafelski:2014fqa}, consistently with the physical
picture underlying this model. Additional support for this picture comes from
the combined study of yields and transverse momentum distributions
discussed in \cite{Begun:2013nga,Begun:2014rsa}. The main weak points
of the non-equilibrium fits come from the additional free parameters
and the relatively small number of particles included in the fits.

In Fig.~\ref{fig:share-different-fits}, SHARE is used both in the
equilibrium and in the non-equilibrium mode (\gq\ fixed to unity or
free) in order to fit the ALICE measurement, including or not nuclei.
It the equilibrium model, nuclei follow the systematics established by
other particles, as discussed above. The non-equilibrium model, on the
other hand, overestimates nuclei production by a large factor. If the
nuclei are included in the fit, the \gq\ parameter converges to unity
and the fit quality degrades, with deviations similar to the
equilibrium fits.

\paragraph{Hadronic interactions}
Inelastic processes in the hadronic phase may not be completely
negligible as normally assumed and could affect in a stronger way
baryons than mesons. In particular, it was suggested that baryons
annihilation in the hadronic phase could have a major role in the
reduction of the proton yields. These effects have been studied
quantitatively in~\cite{Becattini:2014hla,Becattini:2012xb} using the
UrQMD hadron transport code. They have been found to be consistent
with the data. The main criticism to this approach is the fact that
reverse reactions with more than 2 bodies are so far not implemented
in UrQMD, even if it has been argued that they will not suppress the
effect of the annihilation (see also~\cite{Pan:2012ne}).

\paragraph{Flavor hierarchy at freeze-out} Since the transition from
the deconfined medium to hadronic matter is a smooth crossover, the
freeze-out temperature cannot be uniquely identified but it depends on
the observable under study.  It was further suggested
in~\cite{Bellwied:2013cta}, based on lattice and Hadron Resonance Gas
(HRG) calculations, that the (crossover) transition temperature for
light and strange quarks may be different (150~MeV and 165~MeV,
respectively).  This would subsequently lead to a separate freeze-out
at slightly lower temperatures. In this picture, a single temperature
should not describe all data, explaining the tensions observed in the
thermal model fits.  The lattice and HRG calculations can be directly
confronted with the experiments using higher moments of conserved
charges (net charge, net baryon number, net strangeness). Recently,
the STAR collaboration measured the higher order moments of the net
charge and net proton
distribution~\cite{Adamczyk:2014fia,Adamczyk:2013dal} as a function of
beam energy.  These have been compared with theoretical calculations
in~\cite{Borsanyi:2014ewa,Alba:2014eba} to extract a light flavor
freeze-out temperature and baryochemical potential, which were then
used to compute particle ratios in~\cite{Alba:2014eba,Bluhm:QM14}. The
ratios of light flavor particles are found to be consistent with these
parameters, while strange particles are underestimated, consistent
with the expectation. In this comparison there are however important
caveats related to the limitations of the HRG models, to the limited
acceptance of the detectors and to the fact that the experiments
measure net protons and not the real conserved charge (net baryon). A
key crosscheck would be a measurement of net-strangeness, but this may
not be easily attainable as it requires the measurement of several
multistrange baryons.

\paragraph{Outlook} At least four different mechanisms have been
proposed to explain the yields measured at the LHC, leading to a
significant spread in the estimated chemical freeze-out temperature
(\Tch\ in the range 130--165 MeV). It is not yet established which is
the correct mechanisms and how this temperature relates to the QCD
phase transition temperature. 

Additional experimental constraints could help to identify the correct mechanism, in particular:
\begin{itemize}
\item Precise measurements of the trends of particle ratios as a
  function of centrality (to constrain the effect of the hadron phase,
  expected to be strongly centrality dependent; current measurements
  at the LHC do not show significant changes within uncertainties for
  most particle ratios);
\item Measurement of ``heavy'' light-flavor mesons (for instance, the
  expectations for the $\omega$ meson in the equilibrium and
  non-equilibrium models differ by $\sim 1.4$);
\item Measurements at lower energy with the improved vertex detectors
  (this would reduce some uncertainties, for instance feed-down
  uncertainties in the proton measurements);
\item Additional higher moment measurements at LHC and RHIC, in
  particular net strangeness (to constrain the freezeout temperature
  of strange particles in the hierarchical freezeout scenario).
\end{itemize}

\section*{Acknowledgments}
\label{sec:ack}

I am indebted to many colleagues for the stimulating discussion, for
the precious suggestions and for providing me with the experimental
data and theoretical calculations: P.~Alba, F.~Becattini,
J.~Noronha-Hostler, M.~Petr\'an, C.~Ratti, K.~Redlich; the
colleagues from the STAR and HADES collaborations, in particular
L.~Kumar, M.~Lorenz, N.~Xu, X.~Zhu; my colleagues from the ALICE
collaboration, in particular A.~ Andronic, F.~Antinori, R.~Bellwied,
B.~D\"onigus, B.~Hippolyte, A.~Kalweit, H.~Oeschler, M.~Ploskon,
R.~Preghenella, K.~Safarik, J.~Schukraft.





\bibliographystyle{elsarticle-num}
\bibliography{mfloris-qm14-proceedings.bib}

\end{document}